# Cyber-Physical Systems, a new formal paradigm to model redundancy and resiliency

Mario Lezoche, Hervé Panetto

*Research Centre for Automatic Control of Nancy (CRAN), CNRS,*
*Université de Lorraine, UMR 7039, Boulevard des Aiguillettes*
*B.P.70239, 54506 Vandoeuvre-lès-Nancy, France.*
*e-mail: {mario.lezoche, herve.panetto}@univ-lorraine.fr*

Cyber-Physical Systems (CPS) are systems composed by a physical component that is controlled or monitored by a cyber-component, a computer-based algorithm. Advances in CPS technologies and science are enabling capability, adaptability, scalability, resiliency, safety, security, and usability that will far exceed the simple embedded systems of today. CPS technologies are transforming the way people interact with engineered systems. New smart CPS are driving innovation in various sectors such as agriculture, energy, transportation, healthcare, and manufacturing. They are leading the 4-th Industrial Revolution (Industry 4.0) that is having benefits thanks to the high flexibility of production. The Industry 4.0 production paradigm is characterized by high intercommunicating properties of its production elements in all the manufacturing processes. This is the reason it is a core concept how the systems should be structurally optimized to have the adequate level of redundancy to be satisfactorily resilient. This goal can benefit from formal methods well known in various scientific domains such as artificial intelligence. So, the current research concerns the proposal of a CPS meta-model and its instantiation. In this way it lists all kind of relationships that may occur between the CPSs themselves and between their (cyber- and physical-) components. Using the CPS meta-model formalization, with an adaptation of the Formal Concept Analysis (FCA) formal approach, this paper presents a way to optimize the modelling of CPS systems emphasizing their redundancy and their resiliency.

1. **Introduction**

The term cyber-physical systems (CPS) refers to a generation of systems with integrated computational and physical capabilities. Those systems can interact with other systems through many modalities. The interconnected systems create a large range of network that are multi-disciplinary and physically-aware engineered systems, adapted from (Gunes et al., 2014). Inside this kind of network there are components with advanced abilities: sensing, data collection, data transmission and mechanical actuation.

In cyber-physical systems, physical and software components are deeply entangled. The components operate on different spatial and temporal scales interacting with each other in multiple ways that are context related. Opportunities and research challenges include the design and development of next-generation airplanes and space vehicles, hybrid gas-electric vehicles, fully autonomous urban driving, and prostheses that allow brain signals to control physical objects.

The CPS concept is strongly related to the 4-th Industrial Revolution (Industry 4.0). This new industrial conception will have benefits from high flexibility of production, easy and more accessible contribution from all involved entities of business processes (Zuehlke, 2010). The strategic imperatives in this regard reiterated vendors to embed enhanced functionality into the core of enterprise systems to support industry strategy in pursuit of the smart factory for responsive, agile, holonic, and resilient manufacturing. Those functionalities could harness and manage changes and evolutions. Personal interviews with industry experts clearly demanded for strategic alliances with the technology vendors to ensure incremental enhancement in the functions in pursuit of a smart factory to ultimately earn manufacturing excellence and resilience (Asif et al., 2018).

Each component in both physical facilities and cyber devices needs an appropriate level of security guard. A scalable and distributed coordination layer is very important to ensure its resiliency. In Industry 4.0, the CPS potentially involve trillions of devices, and the communication among devices must be highly scalable. The research in this domain proposed several methods to find meaningful information from a large amount of noisy data by conducting trustworthiness linkage inference on a constructed object-alarm graph (Li et al., 2018). To accelerate the assessment of this new conception, the following research directions related to the CPS play a key role: optimization of sensor networks organization, structural aspects for redundancy optimization, resiliency structures. Those fields can benefit from the artificial intelligence research domain. The actual investigation is about the application of an approach named Formal Concept Analysis (FCA) for modelling and analysing a large-scale set of collaborative systems (Lezoche et al., 2012), specifically CPSs.

The research addressed in this paper is related to the study of FCA-based patterns for optimizing CPSs structure related to the properties of **redundancy** and **resiliency**.

In the process of CPS software development (Yiping et al., 2016) stated that one should focus on events more than on functions or objects. In our research we focus on core function determination from system knowledge. Before addressing this topic, we are presenting our definition of what a CPS is, though a meta-model showing all kind of relationships that may occur between CPSs themselves and between their (cyber- and physical-) components.

## 2. A Meta-Model of Cyber-Physical Systems

Components of a CPS: lets denote as Pi and Ci respectively the set of physical and cyber components of a system CPSj. CPSj is a structural agglomerate of these elements Pi and Ci which can also include other subsystems CPSk into a composite cyber-physical system. There are two relations of different nature between these components:

- $R^P$ - the relation between subsystems to be physically connected (e.g. in a production line) and signifies transmission of any kind of physical object between systems.

- $R^C$ - the relation of the connection between cyber components which signifies the presence of an information/control channel between the components.

The components of a system perform certain functions depending on their role in the system, and according to that they have inputs $I_i$ and outputs $O_i$, that capture the flows between this element and the elements that are related to by $R^P$ and $R^C$. As an example, for a sensor, inputs and output reflects the transformation of mechanical or physical alterations of the physical world into a signal that quantitatively measures a particular property. The source and destination of the exchange can be either other components of the system or the environment or some external sources or sinks. To cover the latter case, we introduce into the sets of all physical and all cyber elements of CPSs, a model composed of two elements $e_P$ and $e_C$ standing respectively for these environmental or external sources or destinations.

We define a system of CPSs as a tuple CPSs = $\langle \mathcal{P}, \mathcal{C}, \mathcal{CPS}, R^P, R^C \rangle$, where $\mathcal{P} = e_P \cup \bigcup_i P_i$ is a set of physical components of individual CPS, $\mathcal{C} = e_C \cup \bigcup_i C_i$ is a set of cyber components, and $\mathcal{CPS}$ which is set of CPSs. Each individual CPS of the set $\mathcal{CPS}$, as it was defined before, is a tuple of subset of cyber components, physical components and other CPS that it consolidates. Here we assume that every element of $\mathcal{P}, \mathcal{C},$ and $\mathcal{CPS}$ has its corresponding inputs $I_i$ and outputs $O_i$. The details of exchange do not have special interest for our research in this paper, in general $I_i$ and $O_i$ can be of any type and have any values or contents.

Compositionally, different CPSs $c_1, c_2 \in \mathcal{CPS}$ could share some of their components: $c_1 \cap c_2 \neq \emptyset$. For example, as in systems utilizing the same computational node to supervise physical production activities, or as an actuator such as a light switch which can be considered as a part of two systems: one is the local electrical circuit of an apartment, and the other is a smart-home system for automating and controlling the household electronics. Figure 1 is an example of two simple CPSs consisting of one physical ($P_i$) and one cyber ($C_i$) component each forming a composite CPS. The communication between $CPS_1$ and $CPS_2$ is done through the relations between $C_1$ and $C_2$; and the composite $CPS_3$ has its own actuator components $P_3$ and $C_3$.

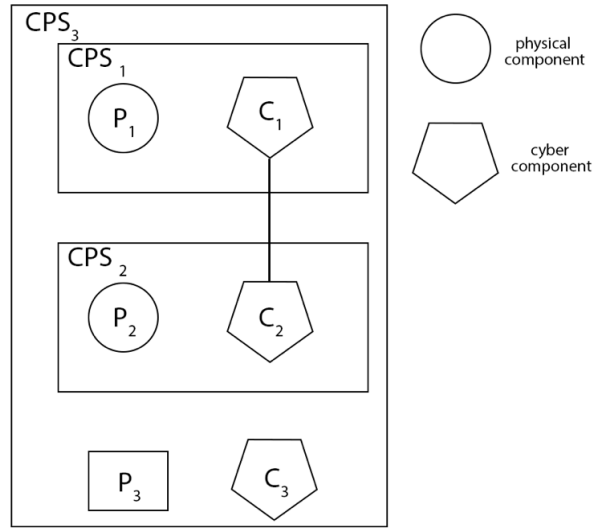

Figure 1, Composite $CPS_3$ consisting of two subordinate systems $CPS_1$ and $CPS_2$ and providing its proper functionality through components $P_3$ and $C_3$.

*2.1 CPS Meta model*

The proposed meta-model CPSs = $\langle \mathcal{P}, \mathcal{C}, \mathcal{CPS}, R^P, R^C \rangle$ that we have elaborated is presented in UML 2.0 notation on physical properties of domains with its spatial structure. Some other authors (Son et al., 2012) show how to transform a Simulink model into an ETRI CPS Modeling Language (ECML) model for modelling CPS for simulating its behaviour and (Klimeš, 2014) shows how to control cyber-physical systems deriving behavioural specifications from user inputs. All these researches focused on the design and the internal behaviour of a CPS. Our work presents a formal meta-model of the structure of any CPS, for proposing a common formal foundation of a composite CPS, aggregating the broad definitions found in the literature.

Figure 2. In the scientific literature, some authors have proposed different results related to CPS meta models from different points of view. (Jeon at al., 2012) present the CPS Meta Modeller tool for designing complex and large-scale systems using the Electronics and Telecommunication Research Institute (ETRI[1]) CPS Modelling Language. (Mezhuyev et al., 2013) show the geometrical meta-metamodel, allowing to link the

---

[1] https://www.etri.re.kr/eng/main/main.etri

The elaborated meta model finds its focus in the interaction of the cyber component, which naturally stands for its computational functionality, and the physical component, which models its physical behaviour. The existence of those two entities let emerge the concept of Cyber-Physical System. If one of those components doesn't exist there is no possibility to have a CPS. The Physical component is modelled as an abstract class, it could be composed by a sensing component, an actuating component or by a component that merges the two capabilities. A CPS component needs an input and an output. It cannot exist a CPS component that has got only one of those two properties. An atomic CPS is the one that does not have any subsystems, but his own functional elements. This definition is created to show, with the best detail possible, the relationships between the two different parts of the entire CPS system. It stops at the presented "atomic" level because of the scope of this scientific paper that focuses on the relationships of the CPSs and the possibility to improve their interoperability.

We also note that we do not make difference between physical and cyber types of communication and corresponding types of input and output interfaces, although it could be a worthwhile extension for a future model.

The relation 'is part of' (physically) is introduced into the model to represent physical structure of systems and their inclusion into one another on the physical level. As an extension to this type of composition of complex CPS we also introduce the aggregation relation 'logically includes'. Together with inheritance relation between classes Composite CPS and Atomic CPS it complies to the structural Composite pattern. With help of this aggregation relation, we model the property of CPS of dynamical reconfiguration and adaptation. Any system can lend its functionality to many super-systems (we borrow the utilisation of sub- and super- prefixes from mathematics, by analogy with subsets and supersets), although probably not at the same time. Inversely, any system can accommodate multiple subsystems.

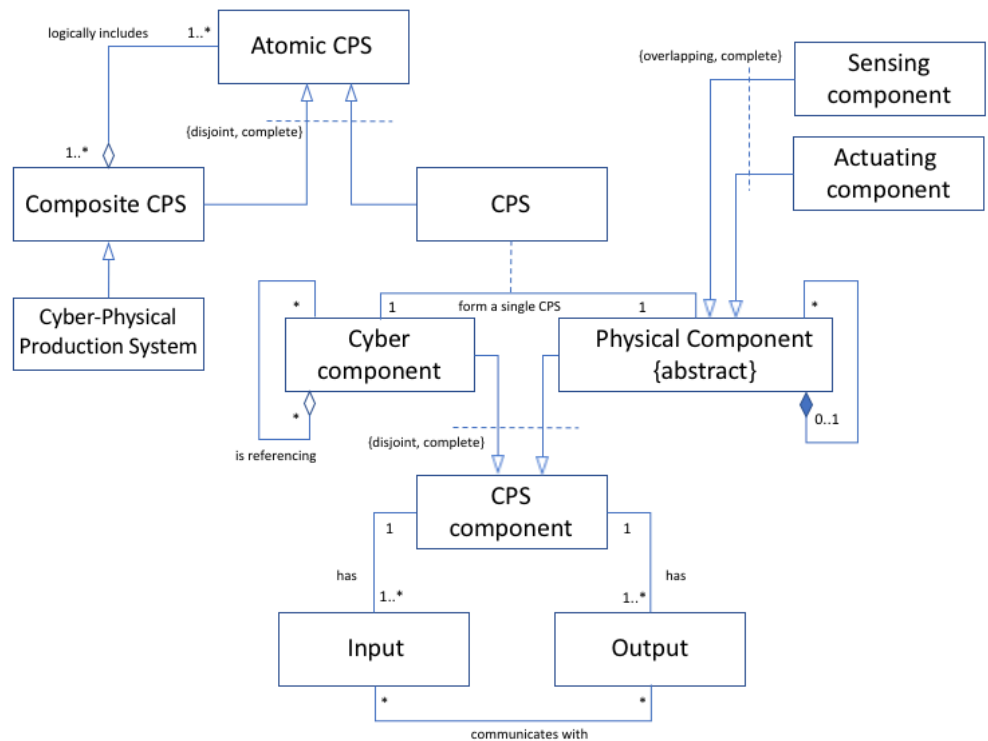

Figure 2 Meta-model of CPS

The class of Cyber-Physical Production Systems can be viewed, in the proposed meta-model, as a subclass of Composite CPS. This interpretation goes in the same direction of the Monostori definition (Monostor, 2014). There is a tight connection between these two relations 'is physically part of' and 'logically includes', in the sense that whenever a system is in the relation 'is physically part of' this also entails that it is being 'logically included' in that system, but not in the other direction.

### 3. Finding Resiliency structure in CPS sub-systems

In the literature, we can find several definitions of the term "resilience". In (Sterbenz et al., 2010) the resilience is the ability to provide and maintain an acceptable level of service in the face of faults and challenges to normal operation. Another definition in (Hollnagel, 2006) positions the resilience as the "ability of a system or organization to react to and recover from disturbances at an early stage with minimal effect on its dynamic stability." There are different ranges of threats and challenges for services that can vary from the simple misconfiguration over large scale natural disasters (Smith, 2011). In order to increase the resilience of a given system, the possible defies and risks need to be recognised and the proper metrics must be defined to protect the endangered service.

Unlike the traditional resiliency method, in which a function is allocated to each abstract component, a CPS is modelled in our work as a set of sub-systems. Each one is represented by a subset of components.

#### *3.1 Related works*

The redundancy is the duplication of critical components or functions of a system, it is the ability of providers to have different alternatives. A resilient system must have duplicate components (physical redundancy) or otherwise more than one configuration for working systems' components (functional redundancy). The higher the degree of redundancy, the more resilient the system would be (Azadeh et al., 2018).

Initially the intention of redundancy is the increasing reliability of the system but it can also present itself as modelling errors. Various methodologies exist in literature to face the redundancy optimization problems. (Aggarwal et al., 1975) state to select the stage where redundancy is to be added, an heuristic criterion is introduced which takes into account the relative increment in reliability versus decrement in performance. (Dokhanchi at al., 2018) propose a framework for the elicitation and debugging of formal specifications for Cyber-Physical Systems through two debugging algorithms. One checks for erroneous or incomplete temporal logic specifications without considering the system and the other can be used for the analysis of reactive requirements with respect to system test traces. The user study establishes that requirement errors are common and that the debugging framework can resolve many insidious specification modellers' errors. The redundancy study is also a way to prevent a single-point of failure, as stated by (Cardenas et al., 2008) or hardware failure that is not considered in design (Wan et al., 2011) thus, practical systems may need to incorporate redundancy. The knowledge of the domain and application context may help to unify information presentation and permits subsystems reuse, so as to reduce information redundancy during the process of semantic interpretation in the agents. (Lin et al., 2010) define an ontology for the system domain to reduce information redundancy in the model and simplify the data interpretation procedure. The redundancy problem has been studied under multiple points of views, our approach focuses the structural evidence coming from the formal knowledge extracted by the data clustering. It has become increasingly crucial that cognitive representations of the models need to be created as a pre-requisite for assessing and intelligently managing the complexity, maneuvering through uncertain environments and eventually achieving the optimized outcomes (Chung-Sheng et al., 2018).

Resilience is not only related with decreasing the probability of failure. It also highlights the need to recuperate from unexpected instabilities in the operating environment (Tierney and Bruneau, 2007). Fundamentally, resilience indicates the ability of a system to recover, it is a function of various system properties, including component

reliability and architecture re-configurability. Resilience can be divided into two categories (Rose, 2007):

(1) the "static resilience" is related to the "ability of an entity or a system to maintain function," or to survive, when disrupted;

(2) the "dynamic resilience" deals with recovery of the system after a shock.

Resilience is highly context dependent — it depends on the architecture of the system, its operational environment, and the disruptive event. For example:

- Different systems are resilient to different disruptions.
- A system could be resilient to one type of disruption but not to another type.

The resilience approaches can also be comprehended through the application ways. Mostly, this refers to using the frameworks for three different purposes, diagnostic, evaluative and planning:

- Diagnostic – evaluating the degree to which a system is resilient to diverse shocks for measuring and quantifying resilience.
- Evaluative – determining the level to which resilience activity is successful in achieving its objective.
- Planning – formulating the consequences of the resilience approach to understand how the planned design fits into the intervention procedure. (Constas et al., 2014).

The Resilience in Industry 4.0 reiterates optimal manufacturing to reap the dream of productivity and manufacturing excellence. Whilst all indications are that ERP system integrates all business processes, yet, the existing enterprise systems lack sophistication for data integration in relation to the manufacturing functionality for online planning and Control (OnP/OnC). Hence, there is a need of holistic research in this domain to optimise the core capabilities of industry (company)-specific production. The implications of enterprise system implementation in the industry 4.0, may act as a precursor for future researchers to contemplate the interaction through whole the product life cycle. As an outcome of the research, the exploration of the role of the big data, the analytics and the cyber-physical systems in achieving the resilience objectives of a smart and connected cyber-physical factory would be a core research topic (Asif et al., 2018).

Most resilience-related research uses some aspect of network theory to study effects of disruptions (Crucitti et al., 2005). Due to the analysis of many systems, it is not possible to assume homogenous nodes. The frameworks that make this assumption are not able to analyse those types of networks. A few studies have considered nodes with the same function but different capacities (Crucitti et al., 2005) and managed to create a roadmap of enabling technologies of CPS in three types of network. Those networks are constructed to manage the resources such as computation power, storage, and bandwidth, and enable resilient services selection. (Chun Kit et al., 2018).

Lately, multilayer networks resilience is acquiring attention as a way to characterise heterogeneous networks. Networks can be modelled as multilayers in two different ways. The network may consist of different physical layers or it may require support from different layers. The results of the studies in this field have stated the concept of interdependent network analyses to highlight the characteristics of such networks (Rinaldi, 2004). Applying these studies to CPSs, designers can study how a failure in one network can have repercussions in the other and how interdependent networks can fail catastrophically after the removal of a small fraction of nodes. These results in turn can guide resource allocation at critical nodes.

Recent research has tried to influence control theory to handle with resilience of interconnected and interdependent systems (Liu, Slotine, and Barabasi 2011) with the ultimate goal of developing resilient controllers (Alessandri and Filippini, 2013).

All the studied frameworks lack the rigorous and formal approach to analyse the structural features. Our proposal focuses on the formal structuration of key features that add more resilience to the studied system.

### *3.2 Definitions*

In this work, we use the following definitions:

**Definition 1 (function):** a function F is a candidate for an abstract component. F is formed by the tuple <I, O>, where I is the set of inputs $I = \{I_1, I_2, ... I_n\}$, and O is the set of outputs $O = \{O_1, O_2, ... O_m\}$.

**Definition 2 (CPS subsystem function):** A CPS subsystem function $ssf \in F$ is an oriented graph $G_{ssf} = (F_{ssf}, E_{ssf})$ which corresponds to a part of a CPS. $F_{ssf} = \{F_{ssf,1}, F_{ssf,2}, ..., F_{ssf,n}\}$ is the set of functions belonging to the sub-system *ssf*, and E is the set of dependency relations that connect these functions, where $<F_{ssf,i}, F_{ssf,j}> = <F_{ssf,i}.O \rightarrow F_{ssf,j}.I>$ denotes the connection between functions $F_{ssf,i}$ and $F_{ssf,j}$ within the subsystem *ssf*.

**Definition 3 (function request):** a function query Q is a directed graph $G_q = (F, E)$, where $S = \{S_1, S_2, ..., S_n\}$ is the set of nodes and each node corresponds to a function requested by a user. E is the set of arcs corresponding to the dependency relations between these requested services.

**Definition 4 (composite function):** a composite function CF is a directed graph $G_f = (SSF, E)$, where SSF is a set of CPS subsystem functions, and E is the set of dependency relations between them. A dependency relation $<ssf_i, ssf_j> = <ssf_i.O \rightarrow ssf_j.I>$ denotes the connection between subsystem functions $ssf_i$ and $ssf_j$ within the candidate composition CF. A composite function CF satisfies a query Q if, and only if, $G_f = (SSF, E)$ and $G_q = (S, E)$ are isomorphic.

## 4. Applying a formal method to structure knowledge model

Formal Concept Analysis (FCA) is a mathematical theory invented by Rudolf Wille in (Wille, 1982). The theory is a bridge between mathematics, data analysis and ordered set theory. This method allows grouping objects according to their common properties (Kumar and Singh, 2014). Two key FCA structures are exploited in the present work, namely formal context and Galois lattice. A formal context is the triplet (O, A, R), where O is a set of objects, A is a set of attributes and R = O×A is the set of binary relations between O and A. Assuming that an object o has the attribute a, the relation R between o and a is denoted by oRa. A formal context can be represented by a matrix (O×A), where the lines and columns respectively correspond to the objects and attributes, and the entries denote the relationships between them, that is, $R_{ij}$ = 1 if the object $O_i$ contains the attribute $A_j$. Otherwise, $R_{ij}$ = 0. Table 2 shows an example of formal context K = (O, A, R), where O is a set of objects O = {$O_1$, $O_2$, $O_3$, $O_4$, $O_5$, $O_6$, $O_7$, $O_8$}, A is a set of attributes A = {$A_1$, $A_2$, $A_3$, $A_4$, $A_5$}, and R is a set of binary relationships between O and A.

| Attribute Object | $A_1$ | $A_2$ | $A_3$ | $A_4$ | $A_5$ |
|---|---|---|---|---|---|
| $O_1$ | 1 | 0 | 1 | 0 | 0 |
| $O_2$ | 0 | 1 | 0 | 1 | 0 |
| $O_3$ | 0 | 0 | 1 | 0 | 1 |
| $O_4$ | 1 | 1 | 1 | 0 | 0 |
| $O_5$ | 0 | 0 | 1 | 0 | 0 |
| $O_6$ | 0 | 0 | 0 | 1 | 1 |
| $O_7$ | 0 | 1 | 1 | 0 | 0 |
| $O_8$ | 1 | 0 | 1 | 1 | 0 |

Table 1 - Example of a formal context

From the above formal context, a set of clusters called "formal concepts" are derived to form a complete lattice structure. A lattice of formal concepts can be represented in the form of an ordered diagram, also called a Hasse diagram. Each node of this diagram corresponds to a cluster and the arcs represent the relations of inclusion between the concepts. Each formal concept has two sections. The Extent section regroups a sub-set of objects sharing the common attributes, whereas the Intent section contains those attributes shared by objects in the Extent part. The figure 3 shows an example of concept lattice corresponding to the formal context in Table 1

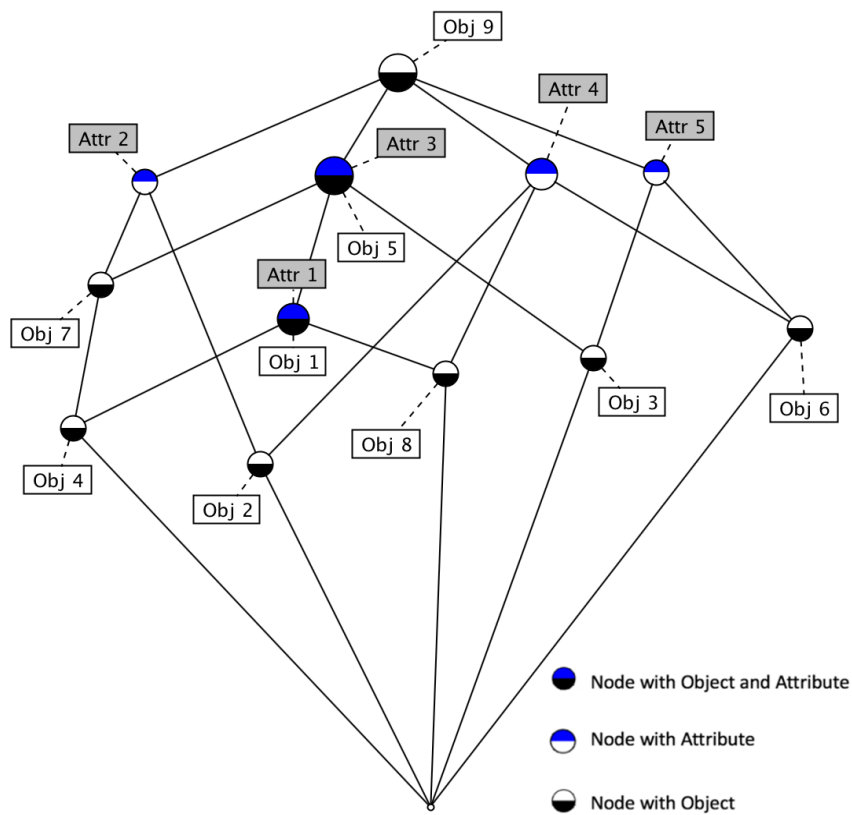

Figure 3 - example of concept lattice

Formal Concept Analysis has been successfully applied in several domains, such as knowledge discovery, road networks modelling, information retrieval, ontology merging, software code analysis, etc. (Poelmans et al., 2013). We propose a method that describes a new way of finding reliable functions by reusing CPS subsystems. To allow the extraction of relevant CPS subsystems, we start in the next section by modelling and clustering the CPS subsystems according to their component functions.

### 5. *Modelling CPS sub-systems functions*

The explicit definition of CPS substitutable functions (CSF), as well as their component functions and the relationships between them, is the first step toward effective composition of CPS function subsystem. In this section, we model such CSF information in the form of a formal context.

Consider the CSF formal context $K^f$ = (SSF, F, R), where F is the set CPS subsystems functions SSF = {$SSF_1$, $SSF_2$, $SSF_3$, ..., $SSF_n$}, F is the set of functions F = {$F_1$, $F_2$, $F_3$, ..., $F_n$} and R is the set of binary relations between fragments and services.

| F<br>SSF | $F_1$ | $F_2$ | $F_3$ | $F_4$ | $F_5$ | $F_6$ |
|---|---|---|---|---|---|---|
| $SSF_1$ | 1 | 1 | 0 | 0 | 0 | 0 |
| $SSF_2$ | 0 | 1 | 1 | 0 | 0 | 0 |
| $SSF_3$ | 0 | 0 | 1 | 1 | 0 | 0 |
| $SSF_4$ | 0 | 0 | 0 | 1 | 1 | 0 |
| $SSF_5$ | 0 | 0 | 0 | 1 | 1 | 1 |
| $SSF_6$ | 0 | 1 | 1 | 1 | 0 | 0 |
| $SSF_7$ | 1 | 1 | 1 | 0 | 0 | 0 |
| $SSF_8$ | 0 | 1 | 0 | 1 | 0 | 1 |

Table 2 shows an example of CSF formal context with eight subsystem functions and six functions that form the user request.

| SSF \ F | F$_1$ | F$_2$ | F$_3$ | F$_4$ | F$_5$ | F$_6$ |
|---|---|---|---|---|---|---|
| SSF$_1$ | 1 | 1 | 0 | 0 | 0 | 0 |
| SSF$_2$ | 0 | 1 | 1 | 0 | 0 | 0 |
| SSF$_3$ | 0 | 0 | 1 | 1 | 0 | 0 |
| SSF$_4$ | 0 | 0 | 0 | 1 | 1 | 0 |
| SSF$_5$ | 0 | 0 | 0 | 1 | 1 | 1 |
| SSF$_6$ | 0 | 1 | 1 | 1 | 0 | 0 |
| SSF$_7$ | 1 | 1 | 1 | 0 | 0 | 0 |
| SSF$_8$ | 0 | 1 | 0 | 1 | 0 | 1 |

Table 2 - Formal context K$^F$ of subsystem functions

As shown in the formal context above, the entries indicate whether a fragment $f \in F$ contains a service s ∈ S (value 1) or s ∉ S (value 0). From the formal context K$^F$, a set of clusters called formal concepts are derived to construct a complete lattice L$_F$ of fragments (see Figure 2). In the literature, there are several lattice construction algorithms such as Bordat, Godin, Dowling (Poelmans et al., 2010). We use the Bordat algorithm (Bordat, 1986) to construct the lattice of fragments. This algorithm is very successful when it runs on contexts of low density. The lattice of fragments is illustrated in Figure 4 and represents the hierarchy of concepts that group the fragments according to their common component services.

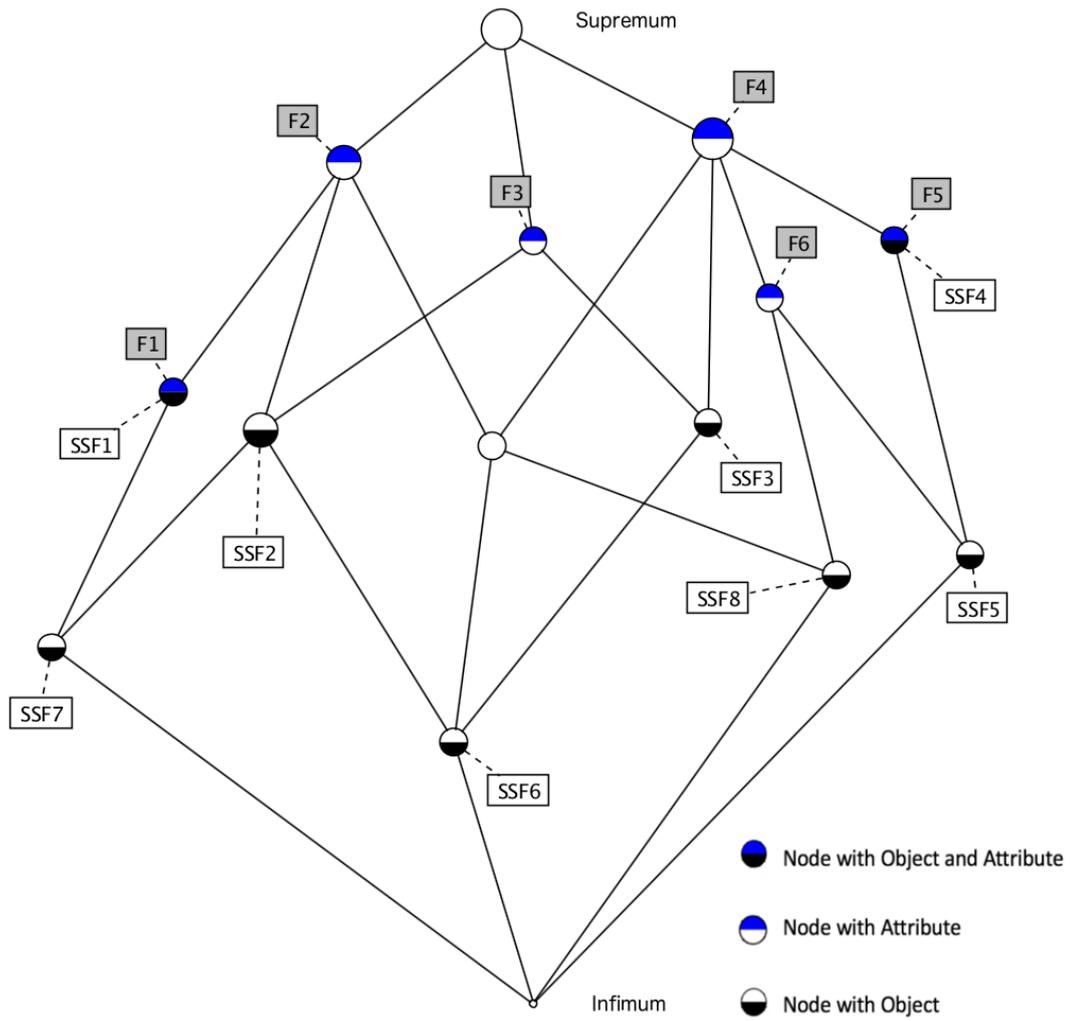

Figure 4 - Lattice $L_F$ of process fragments

The top formal concept in the above lattice is called Supremum, whereas the bottom is called Infimum. From Table 3, we can deduct, from the empty Extent column in the formal concept $C_{15}$, that there is no CPS subsystem offering all the functions {$F_1$, $F_2$, $F_3$, $F_4$, $F_5$, $F_6$} requested by the user.

| Formal Concepts of "CPS Subsystems functions" ||
| Concept Name | < {Concept Extents}, {Concept Intents} > |
| --- | --- |
| $C_1$ | < {$SSF_1$, $SSF_7$}, {$F_1$, $F_2$} > |
| $C_2$ | < {$SSF_2$, $SSF_6$, $SSF_7$}, {$F_2$, $F_3$} > |
| $C_3$ | < {$SSF_3$, $SSF_6$}, {$F_3$, $F_4$} > |
| $C_4$ | < {$SSF_4$, $SSF_5$}, {$F_4$, $F_5$} > |
| $C_5$ | < {$SSF_5$}, {$F_4$, $F_5$, $F_6$} > |
| $C_6$ | < {$SSF_6$}, {$F_2$, $F_3$, $F_4$} > |
| $C_7$ | < {$SSF_7$}, {$F_1$, $F_2$, $F_3$} > |
| $C_8$ | < {$SSF_8$}, {$F_2$, $F_4$, $F_6$} > |
| $C_9$ | < {$SSF_1$, $SSF_2$, $SSF_6$, $SSF_7$, $SSF_8$}, {$F_2$} > |
| $C_{10}$ | < {$SSF_2$, $SSF_3$, $SSF_6$, $SSF_7$}, {$F_3$} > |
| $C_{11}$ | < {$SSF_3$, $SSF_4$, $SSF_5$, $SSF_6$, $SSF_8$}, {$F_4$} > |
| $C_{12}$ | < {$SSF_5$, $SSF_8$}, {$F_4$, $F_6$} > |
| $C_{13}$ | < {$SSF_6$, $SSF_8$}, {$F_2$, $F_4$} > |
| $C_{14}$ | < {$SSF_1$, $SSF_2$, $SSF_3$, $SSF_4$, $SSF_5$, $SSF_6$, $SSF_7$, $SSF_8$}, {} > |
| $C_{15}$ | < {}, {$F_1$, $F_2$, $F_3$, $F_4$, $F_5$, $F_6$} > |

Table 3 - Formal concepts of the context $K^F$

The concept $C_8$ shows, through its Extent column, the existence of a CPS subsystem {$SSF_8$} which offers the functions {$F_2$, $F_4$, $F_6$}. The concept $C_3$ shows that there exist two subsystems {$SSF_3$, $SSF_6$}, which offer two of the requested functions {$F_3$, $F_4$}. Assume that a user sends a request containing the functions {$F_1$, $F_2$, $F_3$, $F_5$}. By running through the above lattice, we can satisfy this request using several combinations of subsystems:

- Combination 1: $C_7$.Extent ∪ $C_4$.Extent

- Combination 2: $C_1$.Extent ∪ $C_2$.Extent ∪ $C_4$.Extent

- Combination 3: $C_1$.Extent ∪ $C_3$.Extent ∪ $C_4$.Extent

Note that the conjunction of several Extent parts may lead to one or more CSF combinations. As an example, the Extent part of the formal concept $C_1$ contains two candidate subsystems ($SSF_1$ and $SSF_7$) that offer two of the five requested functions ($F_1$ and $F_2$), whereas $C_7$.Extent contains only one candidate subsystem ($SSF_7$) that offers three functions ($F_1$, $F_2$ and $F_3$). To satisfy the user's request, some CSF combinations can be extracted from the formal concepts $C_1$, $C_2$ and $C_4$, which are $\{SSF_1, SSF_7\}$, $\{SSF_2, SSF_6\}$ and $\{SSF_4, SSF_5\}$. These CSF combinations are determined by analysing the hidden relations between formal concepts and checking if the function in their Intent parts may satisfy the user needs.

In this way, analysing the formal context through a standard research algorithm, it is easy to highlight the weakness of a system finding the lack of **redundancy** and explicitly the absence of resiliency. The lack of redundancy, as stated also in (Zheng et al, 2017), could expose the whole system to a vulnerability in the systems security domain. The presented method is used to find the absence of redundancy and not to list a gradient of the redundancy level.

## 5.1 Hierarchical structures of the meta-model and the corresponding algebraic lattice representation through a simple case study

Unlike our previous approach (Morozov et al, 2017) where we modelled CPS using Formal Concept Analysis in a standard object-attribute fashion, currently we extend the modelling approach to also take into account links that exist between components and also for hierarchical inclusion of systems one into another according to their composite structure.

In this way, CPSs can be modelled independently in the physical and cyber perspectives using corresponding relations, each one defining an algebraic lattice. The hierarchical structure gives rise to the third lattice that can be used for tacit knowledge recognition and further explicitation.

Given the CPS model in Figure 4, we can list the functions offered by each element of the atomic CPS subsystems. The functionalities offered by the atomic Physical parts of this CPS are resumed in the following

| Physical atomic subsystem | Function offered |
|---|---|
| $P_1$ | $F_{P_1}^1$ |
| $P_2$ | $F_{P_2}^1, F_{P_2}^2$ |
| $P_4$ | $F_{P_4}^1$ |
| $P_5$ | $F_{P_5}^1, F_{P_5}^2$ |
| $P_6$ | $F_{P_6}^1$ |
| $P_7$ | $F_{P_7}^1$ |

Table 4.

Table 4 – Functions offered by Physical part of the CPS atomic subsystems

On the other part the functionalities offered by the atomic Cyber parts of this CPS are resumed in the following

| Cyber atomic subsystem | Function offered |
|---|---|
| $C_1$ | $F_{C_1}^1$ |
| $C_2$ | $F_{C_2}^1, F_{C_2}^2$ |
| $C_4$ | $F_{C_4}^1, F_{C_4}^2$ |
| $C_5$ | $F_{C_5}^1$ |
| $C_6$ | $F_{C_6}^1, F_{C_6}^2$ |
| $C_7$ | $F_{C_7}^1, F_{C_7}^2$ |

Table 5.

Table 5 – Functions offered by Cyber part of the CPS atomic subsystems

At this stage, we will not focus on the Inputs and Outputs of the elementary functions taking in account that, for simplicity, they are the same for the equal function and there is not a partial overlapping of the domains and ranges. Given this information about the CPS functionality we will describe and formalise manually the experts' knowledge that will be a key point to let the model and the formal tool to work properly.

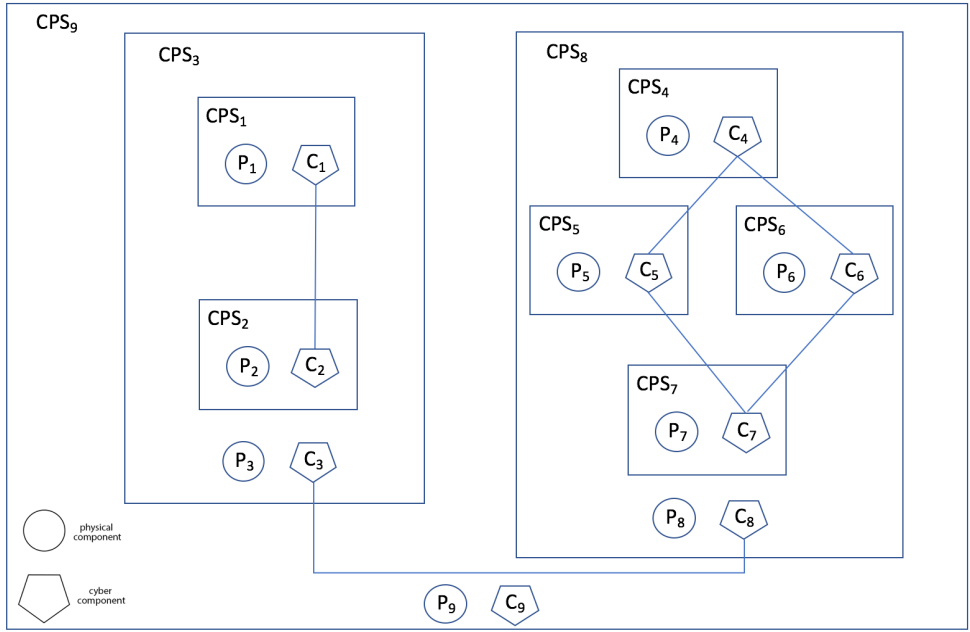

Figure 5 CPS Functional Model example.

In the following

| Physical functions | Cyber functions |
|---|---|
| $F_1^P = F_{P_1}^1 = F_{P_5}^2 = F_{P_7}^1$ | $F_1^C = F_{C_1}^1 = F_{C_7}^2$ |
| $F_2^P = F_{P_2}^1 = F_{P_5}^1$ | $F_2^C = F_{C_2}^2 = F_{C_5}^1 = F_{C_6}^2$ |
| $F_3^P = F_{P_2}^2$ | $F_3^C = F_{C_2}^1 = F_{C_4}^2 = F_{C_6}^1$ |
| $F_4^P = F_{P_4}^1 = F_{P_6}^1$ | $F_4^C = F_{C_4}^1$ |
| - | $F_5^C = F_{C_7}^1$ |

Table 6, we will present what are the same functions offered by the Physical elements and the Cyber elements of the same CPS of Figure 5. We will also name the function to be ready to be used in the formal context creation.

| Physical functions | Cyber functions |
|---|---|
| $F_1^P = F_{P_1}^1 = F_{P_5}^2 = F_{P_7}^1$ | $F_1^C = F_{C_1}^1 = F_{C_7}^2$ |
| $F_2^P = F_{P_2}^1 = F_{P_5}^1$ | $F_2^C = F_{C_2}^2 = F_{C_5}^1 = F_{C_6}^2$ |
| $F_3^P = F_{P_2}^2$ | $F_3^C = F_{C_2}^1 = F_{C_4}^2 = F_{C_6}^1$ |
| $F_4^P = F_{P_4}^1 = F_{P_6}^1$ | $F_4^C = F_{C_4}^1$ |
| - | $F_5^C = F_{C_7}^1$ |

Table 6 – Equal functions offered by Cyber and Physical part of the CPS

In our analysis, we will include also the implicit "part of" property of the atomic subsystems towards the complete one. The following Table 7 presents the three functions representing that knowledge.

| CPS | Inclusive Functions |
|---|---|
| CPS$_3$ = CPS$_1$ + CPS$_2$ | $F_I^1 = \{CPS_1, CPS_2\}$ |
| CPS$_8$ = CPS$_4$ + CPS$_5$ + CPS$_6$ + CPS$_7$ | $F_I^2 = \{CPS_4, CPS_5, CPS_6, CPS_7\}$ |
| CPS$_9$ = CPS$_3$ + CPS$_8$ | $F_I^3 = \{CPS_3, CPS_8\}$ |

Table 7 – Equal functions offered by Cyber and Physical part of the CPS

Taken in account all that information, we can create the formal context for the CPS of Figure 5.

| Functions<br>Subsystems | $F_1^P$ | $F_2^P$ | $F_3^P$ | $F_4^P$ | $F_1^C$ | $F_2^C$ | $F_3^C$ | $F_4^C$ | $F_5^C$ | $F_I^1$ | $F_I^2$ | $F_I^3$ |
|---|---|---|---|---|---|---|---|---|---|---|---|---|
| CPS₁ | 1 | 0 | 0 | 0 | 1 | 0 | 0 | 0 | 0 | 1 | 0 | 1 |
| CPS₂ | 0 | 1 | 1 | 0 | 0 | 1 | 1 | 0 | 0 | 1 | 0 | 1 |
| CPS₄ | 0 | 0 | 0 | 1 | 0 | 0 | 1 | 1 | 0 | 0 | 1 | 1 |
| CPS₅ | 1 | 1 | 0 | 0 | 0 | 1 | 0 | 0 | 0 | 0 | 1 | 1 |
| CPS₆ | 0 | 0 | 0 | 1 | 0 | 1 | 1 | 0 | 0 | 0 | 1 | 1 |
| CPS₇ | 1 | 0 | 0 | 0 | 1 | 0 | 0 | 0 | 1 | 0 | 1 | 1 |

Table 8 – Formal context of Figure 5 CPS

From this formal context, we are able to automatically produce the following formal concepts, summarised in the Table 9.

|  | **Formal Concepts** |
|---|---|
| **Concept Name** | **< {Concept Extents}, {Concept Intents} >** |
| 1 | < {CPS$_1$}, {$F_1^C, F_1^P, F_I^1, F_I^3$}> |
| 2 | < {CPS$_2$}, {$F_2^C, F_3^C, F_2^P, F_3^P, F_I^1, F_I^3$}> |
| 3 | < {CPS$_4$}, {$F_3^C, F_4^C, F_4^P, F_I^2, F_I^3$}> |
| 4 | < {CPS$_5$}, {$F_2^C, F_1^P, F_2^P, F_I^2, F_I^3$}> |
| 5 | < {CPS$_4$, CPS$_6$}, {$F_2^C, F_3^C, F_4^P, F_I^2, F_I^3$}> |
| 6 | < {CPS$_7$}, {$F_1^C, F_5^C, F_1^P, F_I^2, F_I^3$}> |
| 7 | < {CPS$_1$, CPS$_2$}, {$F_I^1, F_I^3$}> |
| 8 | < {CPS$_1$, CPS$_2$, CPS$_4$, CPS$_5$, CPS$_6$, CPS$_7$}, {$F_I^3$}> |
| 9 | < {CPS$_1$, CPS$_5$, CPS$_7$}, {$F_I^1, F_I^3$}> |
| 10 | < {CPS$_1$, CPS$_5$, CPS$_7$}, {$F_1^C, F_1^P, F_I^3$}> |
| 11 | < {CPS$_2$, CPS$_4$, CPS$_6$}, {$F_3^C, F_I^3$}> |
| 12 | < {CPS$_2$, CPS$_5$}, {$F_2^C, F_2^P, F_I^3$}> |
| 13 | < {CPS$_2$, CPS$_4$, CPS$_6$}, {$F_2^C, F_3^C, F_I^3$}> |
| 14 | < {CPS$_4$, CPS$_5$, CPS$_6$, CPS$_7$}, {$F_I^2, F_I^3$}> |
| 15 | < {CPS$_4$, CPS$_6$}, {$F_3^C, F_2^P, F_I^2, F_I^3$}> |
| 16 | < {CPS$_4$, CPS$_5$, CPS$_6$, CPS$_7$}, {$F_2^C, F_I^2, F_I^3$}> |
| 17 | < {CPS$_5$, CPS$_7$}, {$F_1^P, F_I^2, F_I^3$}> |
| 18 | < {CPS$_1$, CPS$_2$, CPS$_4$, CPS$_5$, CPS$_6$, CPS$_7$}, {$F_2^C, F_I^3$}> |
| 19 | < { }, {$F_1^C, F_2^C, F_3^C, F_4^C, F_5^C, F_1^P, F_2^P, F_3^P, F_4^P, F_I^1, F_I^2, F_I^3$}> |

Table 9 – Formal concepts of Figure 5 CPS

Analysing the formal context in

| Functions Subsystems | $F_1^P$ | $F_2^P$ | $F_3^P$ | $F_4^P$ | $F_1^C$ | $F_2^C$ | $F_3^C$ | $F_4^C$ | $F_5^C$ | $F_I^1$ | $F_I^2$ | $F_I^3$ |
|---|---|---|---|---|---|---|---|---|---|---|---|---|
| **CPS$_1$** | 1 | 0 | 0 | 0 | 1 | 0 | 0 | 0 | 0 | 1 | 0 | 1 |
| **CPS$_2$** | 0 | 1 | 1 | 0 | 0 | 1 | 1 | 0 | 0 | 1 | 0 | 1 |
| **CPS$_4$** | 0 | 0 | 0 | 1 | 0 | 0 | 1 | 1 | 0 | 0 | 1 | 1 |
| **CPS$_5$** | 1 | 1 | 0 | 0 | 0 | 1 | 0 | 0 | 0 | 0 | 1 | 1 |
| **CPS$_6$** | 0 | 0 | 0 | 1 | 0 | 1 | 1 | 0 | 0 | 0 | 1 | 1 |
| **CPS$_7$** | 1 | 0 | 0 | 0 | 1 | 0 | 0 | 0 | 1 | 0 | 1 | 1 |

Table 8 and the formal concepts in

| Formal Concepts | |
|---|---|
| Concept Name | < {Concept Extents}, {Concept Intents} > |
| 1 | < {CPS$_1$}, {$F_1^C, F_1^P, F_I^1, F_I^3$}> |
| 2 | < {CPS$_2$}, {$F_2^C, F_3^C, F_2^P, F_3^P, F_I^1, F_I^3$}> |
| 3 | < {CPS$_4$}, {$F_3^C, F_4^C, F_4^P, F_I^2, F_I^3$}> |
| 4 | < {CPS$_5$}, {$F_2^C, F_1^P, F_2^P, F_I^2, F_I^3$}> |
| 5 | < {CPS$_4$, CPS$_6$}, {$F_2^C, F_3^C, F_4^P, F_I^2, F_I^3$}> |
| 6 | < {CPS$_7$}, {$F_1^C, F_5^C, F_1^P, F_I^2, F_I^3$}> |
| 7 | < {CPS$_1$, CPS$_2$}, {$F_I^1, F_I^3$}> |
| 8 | < {CPS$_1$, CPS$_2$, CPS$_4$, CPS$_5$, CPS$_6$, CPS$_7$}, {$F_I^3$}> |
| 9 | < {CPS$_1$, CPS$_5$, CPS$_7$}, {$F_I^1, F_I^3$}> |
| 10 | < {CPS$_1$, CPS$_5$, CPS$_7$}, {$F_1^C, F_1^P, F_I^3$}> |
| 11 | < {CPS$_2$, CPS$_4$, CPS$_6$}, {$F_3^C, F_I^3$}> |
| 12 | < {CPS$_2$, CPS$_5$}, {$F_2^C, F_2^P, F_I^3$}> |
| 13 | < {CPS$_2$, CPS$_4$, CPS$_6$}, {$F_2^C, F_3^C, F_I^3$}> |
| 14 | < {CPS$_4$, CPS$_5$, CPS$_6$, CPS$_7$}, {$F_I^2, F_I^3$}> |
| 15 | < {CPS$_4$, CPS$_6$}, {$F_3^C, F_2^P, F_I^2, F_I^3$}> |
| 16 | < {CPS$_4$, CPS$_5$, CPS$_6$, CPS$_7$}, {$F_2^C, F_I^2, F_I^3$}> |
| 17 | < {CPS$_5$, CPS$_7$}, {$F_1^P, F_I^2, F_I^3$}> |
| 18 | < {CPS$_1$, CPS$_2$, CPS$_4$, CPS$_5$, CPS$_6$, CPS$_7$}, {$F_2^C, F_I^3$}> |
| 19 | < { }, {$F_1^C, F_2^C, F_3^C, F_4^C, F_5^C, F_1^P, F_2^P, F_3^P, F_4^P, F_I^1, F_I^2, F_I^3$}> |

Table 9, we can easily produce lattices representing various information of the studied CPS and following the CPS Meta-model presented in Figure 2.

The lattices in Figure 6 represents the Cyber part (6a) and the Physical part (6b) of the CPS presented in Figure 5.

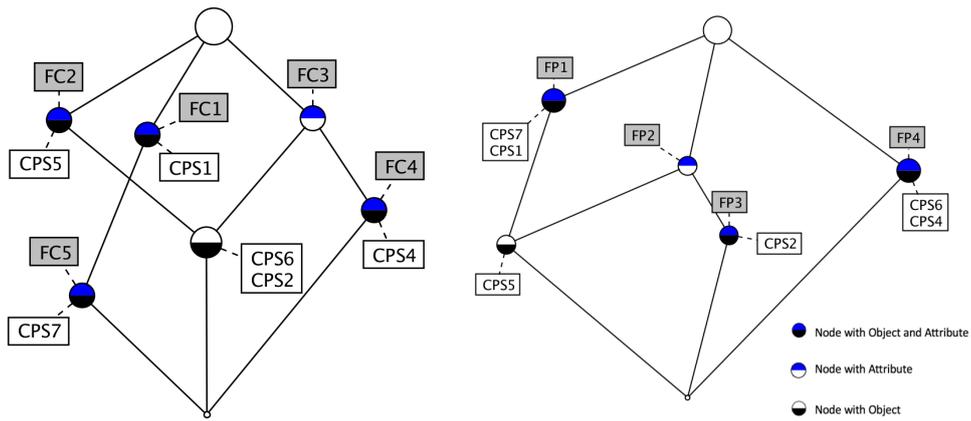

Figure 6 – 6a: Cyber part of CPS – 6b: Physical part of CPS

The lattice in Fig 7 represents the CPS in its completeness highlighting the "part of" relations through the inclusive functions $F_I^1$, $F_I^2$, $F_I^3$.

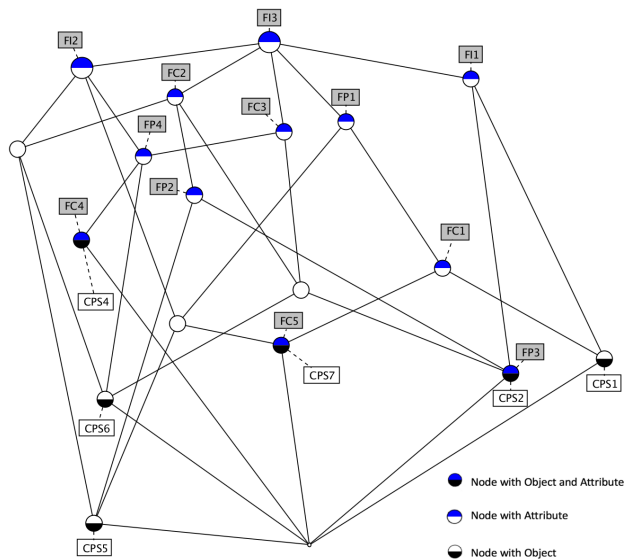

Figure 7 – Lattice representation of CPS depicted on Figure 5

As we can simply deduce, from the

| Formal Concepts | |
|---|---|
| Concept Name | < {Concept Extents}, {Concept Intents} > |
| 1 | < {CPS$_1$}, {$F_1^C, F_1^P, F_I^1, F_I^3$}> |
| 2 | < {CPS$_2$}, {$F_2^C, F_3^C, F_2^P, F_3^P, F_I^1, F_I^3$}> |
| 3 | < {CPS$_4$}, {$F_3^C, F_4^C, F_4^P, F_I^2, F_I^3$}> |
| 4 | < {CPS$_5$}, {$F_2^C, F_1^P, F_2^P, F_I^2, F_I^3$}> |
| 5 | < {CPS$_4$, CPS$_6$}, {$F_2^C, F_3^C, F_4^P, F_I^2, F_I^3$}> |
| 6 | < {CPS$_7$}, {$F_1^C, F_5^C, F_1^P, F_I^2, F_I^3$}> |
| 7 | < {CPS$_1$, CPS$_2$}, {$F_I^1, F_I^3$}> |
| 8 | < {CPS$_1$, CPS$_2$, CPS$_4$, CPS$_5$, CPS$_6$, CPS$_7$}, {$F_I^3$}> |
| 9 | < {CPS$_1$, CPS$_5$, CPS$_7$}, {$F_I^1, F_I^3$}> |
| 10 | < {CPS$_1$, CPS$_5$, CPS$_7$}, {$F_1^C, F_1^P, F_I^3$}> |
| 11 | < {CPS$_2$, CPS$_4$, CPS$_6$}, {$F_3^C, F_I^3$}> |
| 12 | < {CPS$_2$, CPS$_5$}, {$F_2^C, F_2^P, F_I^3$}> |
| 13 | < {CPS$_2$, CPS$_4$, CPS$_6$}, {$F_2^C, F_3^C, F_I^3$}> |
| 14 | < {CPS$_4$, CPS$_5$, CPS$_6$, CPS$_7$}, {$F_I^2, F_I^3$}> |
| 15 | < {CPS$_4$, CPS$_6$}, {$F_3^C, F_2^P, F_I^2, F_I^3$}> |
| 16 | < {CPS$_4$, CPS$_5$, CPS$_6$, CPS$_7$}, {$F_2^C, F_I^2, F_I^3$}> |
| 17 | < {CPS$_5$, CPS$_7$}, {$F_1^P, F_I^2, F_I^3$}> |
| 18 | < {CPS$_1$, CPS$_2$, CPS$_4$, CPS$_5$, CPS$_6$, CPS$_7$}, {$F_2^C, F_I^3$}> |
| 19 | < {}, {$F_1^C, F_2^C, F_3^C, F_4^C, F_5^C, F_1^P, F_2^P, F_3^P, F_4^P, F_I^1, F_I^2, F_I^3$}> |

Table 9, the Figure 6 and the Figure 7 lattices, that the analysed CPS is not fully resilient. In fact, there are three functions that are nowhere duplicated. If the following functions $F_3^P$, $F_4^C$ and $F_5^C$ would stop working no other CPS subsystem could replace them. Using the Figure 6 representation we can rapidly identify the CPS redundancy focusing on the node labels. The node that includes more than one CPS subsystems shows the existence of the redundancy subsystem in the entire system. As an example, in the Figure 6a we find the CPS$_2$ and CPS$_6$ duplicating the same cyber functions and in the Figure 6b we find the CPS$_4$ and CPS$_6$ duplicating the same physical functions.

### 5.2 A simple real case study to present the proposed process

In this section we schematize an enterprise with four production lines and we will show the process to perform a redundancy and a resiliency analysis.

The Figure 8 presents the schema of the shop floor where the four manufacturing lines are controlled by two cameras that send the data flow towards a malfunctions detection software. Each production line is composed by a machine, that performs a particular manufacturing function FRx, linked to some intelligent sensors measuring some physical dimensions. The following definitions explain the various functions:

- FWx represent the water sensors monitoring if there is unattended humidity in parts of the production line where it should not be.
- FPx represent the pressure sensors monitoring a level of pressure not tolerable by the manufacturing machines.
- FTx represent the temperature sensors monitoring the environmental temperature.
- FCx represent the camera controlling the four manufacturing lines. The data flow is sent to a software to detect potential malfunctions.

In Table 10 are presented the functions offered by the system. This table represents the experts knowledge related to the presented system.

| Equal Function offered |
|:---:|
| FC = FC1 = FC2 |
| FRa = FR1 = FR4 |
| FRb = FR2 = FR3 |
| FT = FT1 = FT2 |

Table 10 – Equal functions offered by the CPSx

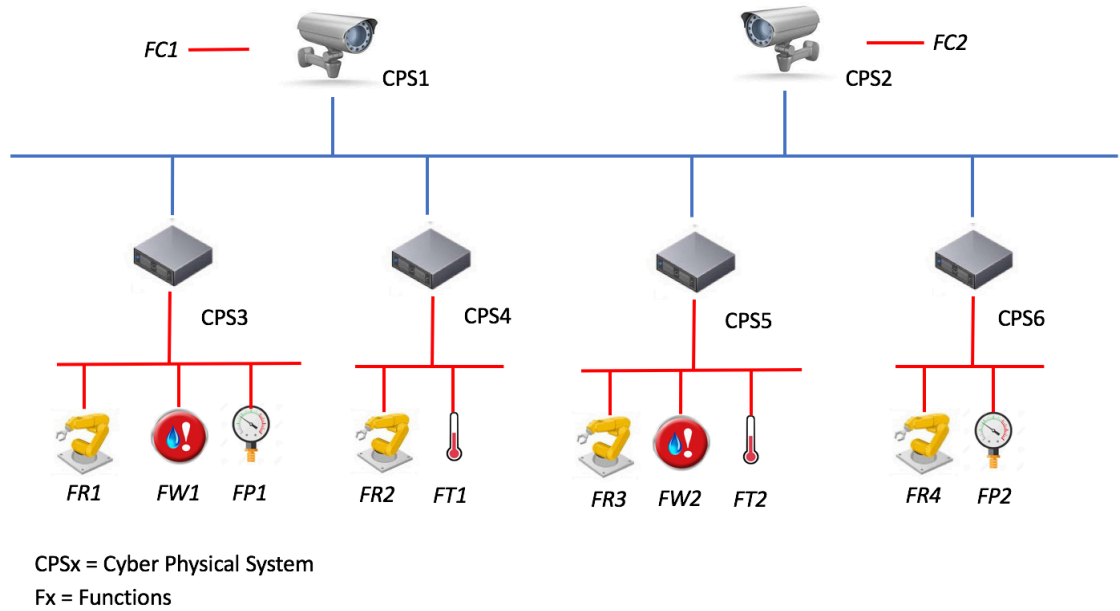

Figure 8 – Four production lines controlled by two cameras

Taken in account all the information from the Table 10 and the Figure 8, we can create the Table 11 in which there will be the formal context for the entire production system.

| Functions<br><br>Subsystems | FC | FRa | FRb | FW1 | FW2 | FP1 | FP2 | FT |
|---|---|---|---|---|---|---|---|---|
| **CPS1** | 1 | 0 | 0 | 0 | 0 | 0 | 0 | 0 |
| **CPS2** | 1 | 0 | 0 | 0 | 0 | 0 | 0 | 0 |
| **CPS3** | 0 | 1 | 0 | 1 | 0 | 1 | 0 | 0 |
| **CPS4** | 0 | 0 | 1 | 0 | 0 | 0 | 0 | 1 |
| **CPS5** | 0 | 0 | 1 | 0 | 1 | 0 | 0 | 1 |
| **CPS6** | 0 | 1 | 0 | 0 | 0 | 0 | 1 | 0 |

Table 11 in which there will be the formal context for the entire production system

From this formal context, we are able to automatically produce the following formal concepts, summarised in the Table 12.

| Formal Concepts | |
|---|---|
| **Concept Name** | **< {Concept Extent}, {Concept Intent} >** |
| 1 | < {CPS1, CPS2}, {FC}> |
| 2 | < {CPS3}, {FRa, FW1, FP1} > |
| 3 | < {CPS4, CPS5}, {FRb, FT} > |
| 4 | < {CPS5}, {FRb, FW2, FT} > |
| 5 | < {CPS6}, {FRa, FP2} > |
| 6 | < {CPS3, CPS6}, {FRa} > |
| 7 | < {CPS1, CPS2, CPS3, CPS4, CPS5, CPS6}, {} > |
| 8 | < {}, {FC, FRa, FRb, FW1, FW2, FP1, FP2, FT} > |

Table 12 Formal concepts derived from the Formal context presented in Table 11.

Given the Formal concepts it is immediate, through a simple research algorithm, finding particular configurations of the intents, the offered functions, that are related to the extents, the subsystems composed by the CPSs.

For examples the controller function, FC, is assured only by the CPS1 and CPS2.

CPS3 and CPS6 have the common manufacturing function FRa but they differ in relation to the other sensors, so a possible logical derivation could be to extend the specific sensors to one another.

The formal concept number 7 gives the information that there is no function offered by all the subsystems.

The formal concept number 8, in a complementary way, states that there is no subsystem that offers all the functionalities.

### 5.3 Assisting the Cyber-Physical Systems Modelling Process

Following the various proposals in literature (Khaitan, 2015), we describe the generalized CPS model as a system of components. Those components can be unmistakeably divided onto two disjoint groups:

1. a control decision and a sensor part, that represents the system cyber layer,
2. a physical counter-part, i.e. all actuators that bridge the information into real-world actions.

This explicit separation between the two system abstracts parts imposes a limitation on the modelling approach. However, its application does not lead to any restrictions on how the system will be later specified or implemented. In our modelling approach, we propose this two layers' division as a separation between the system components functional roles. We consider the physical nodes as terminal execution nodes which highlight the behaviour of the system. In contrast sensor and computation nodes from the cyber layer provide data and decisions. The cases where a cyber-node itself realizes tail end functions can be inferred from the above case by separating its modelling element into two disjoint elements: the physical one which takes over the functions and the cyber one which serves for computations.

Our research investigates the properties of concept lattices, it proposes an illustrative case study of its application to CPS. The comprehension of a system is an iterative process, spacing from a general view to focus towards specific subsystems. In FCA toolset, which is built around a complete lattice, this topic is addressed by arising clusters which covers all levels of generalization. The lattice diagram helps in visual navigation and the implication outlines the axioms proper to the studied domain.

Another relevant topic in the CPSs field is related to deal with big scale systems and the production of a great quantity of data. The literature on FCA proposes various techniques tackling this issue: iceberg concept lattices (Stumme, 2002), projections of pattern structures (Ganther, 2001), and feature selection methods.

## 6. Conclusion

Two results are achieved in this scientific work.

1. The first one is the presentation of a CPS meta model that represents a first step towards the knowledge formalisation needed to structure the use of formal tools,
2. and the second is the FCA adaptation to this domain to find hidden knowledge thanks to the implicit relations existing in the structure of the system.

We analysed and discovered in an automatic and simple way the lack of resilience and the presence of redundancy in the analysed CPS looking directly at the CPS model. The simple example presented show how to focus the weakness of a system and how-to strength its resiliency analysing the Lattice and the formal concepts created from the system model. In the same way, we can balance the redundancy of the system lowering the number of the repeated functions offered by different subsystems. The system context is a key knowledge to optimize the redundancy level. The context expresses reality constraints. Formalizing those constraints and integrating them in the FCA formal context will improve the accuracy of the model and let understand the hidden links between existing concepts. Generally, those links represent the redundant concepts to optimize. This analysis is under study to use the appropriated formal way.

Formal Concept Analysis has been applied in many domains as a knowledge representation and discovery tool. The current paper adapts this approach and its evaluation for the needs of CPS modelling and analysis. The proposed result highlights the FCA bottom-up approach focusing with the particularities of the domain and building upon them a structure to allow to discover the general dependencies. Some research studying relations between the FCA and the graph modelling methods (Carbonnel et al.,

2016), (Morozov et al., 2017) specifies the need of the use of extra filtering after the lattice building process. As a further development of the approach, we plan to study the use of the resulting lattice-models to response some of the pressing questions of Industry 4.0 such as the identification of redundancies in functionalities, the improvement of the systems plasticity and their auto-adaptation to environment changes. The future research would be to implement an approach of reinforcement learning algorithms in the CPS so that, by themselves, following several iterations, the systems could learn the specificities of each other and build, autonomously and in a collaborative way, their own global model. Another feature to add to the presented approach would be the interpretation of the models through some extra knowledge derived from formal representations (e.g. Ontologies). This property would improve the automatic modelling and discovery steps. The formal knowledge representation would be complementary in relation to the informal (expert experiences) source of knowledge. The definition of correspondences between the lattices will be used for a much more optimised improvement of systems' resilience.